\UseRawInputEncoding
\documentclass{pasj01}
\usepackage{graphics}
\draft
\Received{}
\Accepted{}
 
\usepackage{ulem}

\begin{document} 
\title{Model of a `Warm Corona' as the Origin of the Soft X-ray Excess of Active Galactic Nuclei}

\author{Norita \textsc{Kawanaka}\altaffilmark{1,2,3}}%
\altaffiltext{1}{National Astronomical Observatory of
Japan (NAOJ), 2-21-1, Osawa, Mitaka, Tokyo 181-8588,
Japan}
\altaffiltext{2}{Department of Physics, Graduate School of Science Tokyo Metropolitan University 1-1,
Minami-Osawa, Hachioji-shi, Tokyo 192-0397}
\altaffiltext{3}{Center for Gravitational Physics and Quantum Information, Yukawa Institute for Theoretical Physics, Kyoto University, Kitashirakawa Oiwake-cho, Sakyo-ku, Kyoto 606-8502, Japan}
\email{norita@tmu.ac.jp}

\author{Shin \textsc{Mineshige}\altaffilmark{4}}
\altaffiltext{4}{Department of Astronomy, Graduate School of Science, Kyoto University, Kitashirakawa Oiwake-cho, Sakyo-ku, Kyoto 606-8502, Japan}


\maketitle

\begin{abstract}
The soft X-ray excess in the spectra of active galactic nuclei is characterized by similar electron temperatures of 0.1 -- 0.3 keV
and similar photon indices around 2.2 -- 3, if fitted with inverse Comptonization. It remains a puzzle why both values are not sensitive to the black hole mass nor accretion rate. 
Supposing that the scattering-dominated surface layer of an accretion disk 
can act as a warm corona, we construct a vertically one-zone model to understand what determines its temperature.  
By solving the equations of (1) the condition for the effective optical depth,
(2) the energy balance, and (3) dominance of the Compton cooling over the bound-free cooling,
we could reproduce the basic observational features of the soft excess,
provided that anomalous heating (excess heating other than what is expected by local energy dissipation) takes place in the warm corona.
The similar temperatures can be understood, since both of the anomalous heating and Compton cooling rates are proportional to the dissipation rate of the accretion energy, while similar photon indices are a natural consequence of the fact that observed photons are finally emitted from the layer of Compton $y\sim 1$.  The soft excess is not observed in black hole binaries,
since disk temperatures are too high for the Compton scattering to work as cooling.
The derived temperatures are somewhat underestimation, however.
This may indicate a necessity of multi-zone corona structure.  The stability of the warm corona and its consequences are briefly discussed.
\end{abstract}

\section{Introduction}
The enormous amount of radiation energy of an active galactic nucleus (AGN) originates from an accretion disk surrounding a supermassive black hole (SMBH; $M_{\rm BH}\sim 10^{6-9}M_{\odot}$) located at the center of the host galaxy.  The X-ray spectrum of an AGN is primarily characterized by the power-law continuum emission that dominates above $2~{\rm keV}$.  Generally this component is interpreted as the thermal Comptonization of soft photons from an accretion disk by hot electrons in a hot plasma (`corona') whose temperature and scattering optical depth are $T\sim 10^9~{\rm K}$ and $\tau_{\rm es}\sim 1$, respectively \citep{1991ApJ...380L..51H, 1993ApJ...413..507H, 1995ApJ...449L..13S}.  When extrapolating this power-law fit below $\sim 2~{\rm keV}$, however, a significant fraction of type I AGNs show an excess emission in $\sim 0.1-1.0~{\rm keV}$, which is so-called soft X-ray excess \citep{1981ApJ...251..501P, 1984ApJ...281...90H, 1985ApJ...297..633S, 1985MNRAS.217..105A, 1989MNRAS.240..833T, 1993A&A...274..105W, 2004MNRAS.352..523P}.  This component has often been fitted with a blackbody emission, and found that its characteristic temperature is remarkably constant (typically $\sim 0.1-0.3~{\rm keV}$) over a wide range of black hole masses and disk accretion rates \citep{1993A&A...274..105W, 2003A&A...412..317C, 2004MNRAS.349L...7G, 2006MNRAS.365.1067C, 2009A&A...495..421B}.

Two different scenarios have been intensely discussed as the origin of soft X-ray excess: one is the relativistically blurred ionized reflection \citep{2004MNRAS.349L...7G, 2006MNRAS.365.1067C, 2008MNRAS.391.2003Z, 2013MNRAS.428.2901W}, and the other is the Comptonization in a warm corona (in the range of $\sim 0.1-0.5~{\rm keV}$; \cite{1998MNRAS.301..179M, 2011PASJ...63S.925N, 2011A&A...534A..36K, 2012MNRAS.420.1848D, 2013A&A...549A..73P, 2015A&A...580A..77R, 2018A&A...611A..59P, 2020A&A...634A..85P}).
In the present study, we focus on the warm corona model as a plausible explanation for the soft X-ray excess.
The main reason is that the variations of the soft X-ray excess component do not follow those of the hard X-ray components
but those of the optical-UV component (e.g., \cite{2011PASJ...63S.925N, 2011A&A...534A..39M, 2013PASJ...65....4N, 2013A&A...549A..73P, 2018MNRAS.480.3898N, 2020MNRAS.491..532G}).
This is difficult to understand in other models, such as the blurred absorption model and the reflection model,
since then the soft X-ray excess should vary in response to the variations of the hard X-ray component.
Another supporting evidence is that the soft X-ray excess spectra look so smooth (without atomic features)
and can well be fitted with power-law like spectra (e.g., \cite{2020A&A...640A..99M, 2021ApJ...913...13X}), which is also difficult to explain with ionized reflection unless huge blurring is assumed (e.g., \cite{2014A&A...567A..44B}).  The smoothness of soft X-ray spectra is to be examined in more details by future fine-spectroscopic observations such as XRISM.

We should be aware of a number of issues, however, which need to be answered in the framework of the warm corona model:
\begin{itemize}
\item
Similar electron temperatures are reported from the spectral fitting analyses, irrespective of black hole masses $M_{\rm BH}$ or disk accretion rates ${\dot M}$. 
This is not easy to understand in analogue with the standard-type accretion 
disk theory, since the disk temperature is well known to depend on $M_{\rm BH}$ and ${\dot M}$. 
\item
A related question is; what mechanism determines the temperatures of the warm corona? 
In other words, why is there no corona with intermediate temperatures (1 -- 10 keV) observed (except for the important exceptional case of super-Eddington flows)?
\item
Why does the soft excess component exhibit similar photon indices (e.g., \cite{2013A&A...549A..73P}), 
not critically depending on $M_{\rm BH}$ nor ${\dot M}$?
This in turn indicates similar values of the Compton $y$-parameter of the warm corona.
\item
Why is no soft excess observed in stellar-mass black holes?
\item
What determines the strengths of the soft X-ray excess? 
Apparently, it does not depend on $M_{\rm BH}$ nor ${\dot M}$ (e.g., \cite{2004MNRAS.349L...7G}).

\end{itemize}

With these issues kept in mind, we construct a vertically one-zone model to understand what determines the temperature of the scattering-dominated warm corona.
The plan of the present paper is as follows: we first give basic considerations in section 2
so as to understand the basic physical processes underlying the formation of a warm corona.
We then build up a one-zone model in section 3 to see to what extent our model will be able to explain the basic properties of 
the warm corona mentioned above.
Discussion on several theoretical issues, such as the comparison between the bound-free and Compton cooling, thermal stability, 
condition for the anomalous heating in the corona, mass and accretion-rate dependence of the coronal temperature,
will be given in section 4. 
The final section is devoted to concluding remarks.

\section{Basic Considerations}
In this section, we give basic considerations regarding the warm corona structure before constructing a more realistic corona model. 

\subsection{Basic assumptions}
\begin{figure}[ht]
\centering
\includegraphics[width=12.0cm]{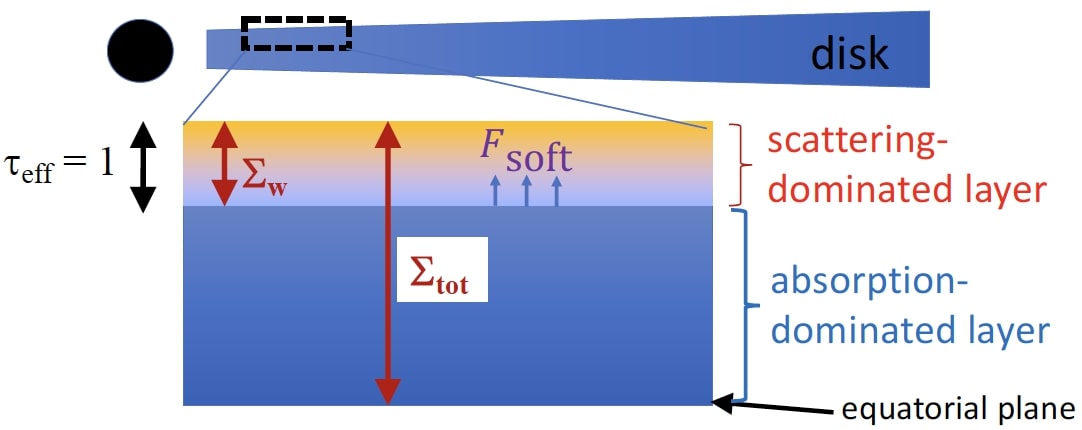}
\caption{A warm corona exposed to soft photon fields (with flux, $F_{\rm soft}$) originating from the disk body.}
\label{fig1}
\end{figure}

The situation which we have in mind is illustrated in Figure \ref{fig1}.  
In general, the thermal soft photons from an optically-thick disk are supposed to emerge from the photosphere, which is defined as the surface where the effective optical depth measured from the infinity is equal to unity: 
\begin{equation}
\tau_{\rm eff}\equiv \sqrt{3\tau_{\rm abs}(\tau_{\rm abs}+\tau_{\rm es})}=1,
\label{tau_eff}
\end{equation}
where $\tau_{\rm abs}$ and $\tau_{\rm es}$ are the absorption optical depth (sum of bound-free and free-free absorption optical depth) and the optical depth with respect to electron scattering, respectively (see \cite{1979rpa..book.....R}, p.50). 
The radiation transfer in the layer above the photosphere is dominated by Compton scattering, while that in the layer below the photosphere is dominated by absorption and re-emission (i.e., reprocessed emission from the layer above). 
The former layer is exposed to copious soft photons (with radiation flux of $F_{\rm soft}$) supplied from the latter layer. 
Our hypothesis is that the former layer (i.e., scattering-dominated layer) can act as a warm corona that generates soft X-ray excess emission. 
Here we denote $\Sigma_{\rm w}$ and $\Sigma_{\rm tot}$ as surface density (above the equatorial plane) of the warm corona and that of the entire disk, respectively.

We study the thermal structure of such a scattering-dominated layer (warm corona) under the following assumptions.
\begin{enumerate}
\item
For simplicity, we adopt a one-zone model for the warm corona structure; 
that is, we assume uniform layer and describe the physical situation of the warm corona by representative physical values, such as temperature, $T_{\rm w}$, and so on. 
\item 
We assume that a fraction $f_{\rm w} (<1)$ of the accretion energy is dissipated within the corona \citep{1991ApJ...380L..51H}. 
We can then write the coronal heating rate (per unit surface area) as
\begin{equation}
  Q_{\rm cor}^{+}= f_{\rm w} Q_{\rm tot}^{+},
\label{corona-heating}
\end{equation}
where $Q_{\rm tot}^+$ represents the total heating rate (per unit surface area)
of the disk-corona system,
\begin{equation}
   Q^+_{\rm tot} = \frac{3}{8\pi} \frac{GM_{\rm BH}{\dot M}}{r^3} f_1,
\label{Qtot}
\end{equation}
where $M_{\rm BH}$, ${\dot M}$, and $r$ are the black hole mass, the mass accretion rate, and the distance from the origin (black hole), respectively,
and $f_1 \equiv 1-\sqrt{r_*/r}$ represents the boundary term (with $r_*$ being the radius of the inner boundary).
Note that if we set equation $Q^+_{\rm tot}=\sigma T_{\rm SSdisk}^4$ (with $\sigma$ being the Stefan-Boltzmann constant, and the subscript ``SSdisk'' denoting the ``Shakura-Sunyaev disk''), we can estimate the surface temperature of the standard-type disks;
\begin{equation}
T_{\rm SSdisk} = 1.1 \times 10^5~{\rm K}~f_1^{-1/4} 
  \left(\frac{M_{\rm BH}}{10^7~M_\odot}\right)^{-1/4} 
  {\dot m}^{1/4}
  \left(\frac{r}{10~ r_{\rm S}}\right)^{-3/4}.
 \label{SSdisk}
\end{equation}
Here, the normalized accretion rate is
${\dot m} \equiv {\dot M}c^2/L_{\rm E}$
with $L_{\rm E}$ being the Eddington luminosity, $c$ being the speed of light, 
and $r_{\rm S} \equiv 2GM_{\rm BH}/c^2$ is the Schwarzschild radius.

\item 
We assume that a half of the coronal emission escapes the warm corona upward, while the other half goes backward and enters the absorption-dominated layer (hereafter, we call as ``the disk body''), thereby heating up there.  We can thus write the energy gain rate within the disk body as
\begin{equation}
  Q_{\rm disk}^{+}= (1-f_{\rm w}) Q_{\rm tot}^{+} 
     + (f_{\rm w}/2) Q_{\rm tot}^{+}  
\label{disk-heating}
\end{equation}
The first and second terms on the right-hand-side represent the energy dissipation within the disk and the incident energy flux from a corona, respectively.

\item
The heated disk body then re-emits soft photons.
If we denote the surface temperature of the disk by $T_{\rm soft}$, 
we have 
$F_{\rm soft} = \sigma T_{\rm soft}^4$,
where we assume blackbody emission.  The disk cooling rate is then
\begin{equation}
  Q_{\rm disk}^- 
  = F_{\rm soft}.
\label{disk-cooling}\end{equation}

\item
The main cooling mechanism of the warm corona is inverse Compton cooling by the soft photons 
emitted from the surface of the disk body. The cooling rate of the corona is thus approximately 
\begin{equation}
　Q^-_{\rm cor} =\frac{4 kT_{\rm w}}{m_{\rm e}c^2} F_{\rm soft}\tau_{\rm es,w}^2,
\label{corona-cooling}
\end{equation}
where $m_{\rm e}$ and $k$ are, respectively, the electron mass 
and the Boltzmann constant, whereas
$T_{\rm w}$ and $\tau_{\rm es,w}$ denote the temperature and the electron-scattering depth of the warm corona, respectively.  
Here we assume that $\tau_{\rm es,w}$ is larger than unity, which is inferred from the observations of soft X-ray excess.
We may rewrite this equation as
\begin{equation}
　Q^-_{\rm cor} = y F_{\rm soft},
\label{corona-cooling2}
\end{equation}
where $y=4kT_{\rm w}/(m_e c^2)\cdot \tau^2_{\rm es,w}$ is the Compton $y$-parameter in the non-relativistic regime \citep{1979rpa..book.....R}.
It is important to note that the Compton $y$-parameter is related to the photon index 
of X-ray spectra through \citep{1999ASPC..161..295B},
\begin{equation}
   \Gamma \approx \frac{9}{4} y^{-2/9}. \label{B99}
\label{Gamma}
\end{equation}
The observed photon indices of $\Gamma \approx 2.2 - 3.2$ (see, e.g., \cite{2018A&A...611A..59P})
leads to $y\approx 0.2 - 1.1$, if we adopt this fitting formula.  \footnote{Note that the equation (\ref{B99}) is evaluated under the spherical geometry, while we here assume the slab geometry in our model.  As long as we adopt a one-zone and one-dimensional model, however, such a distinction will not cause large errors; i.e., rough discussion may be enough. Furthermore, we compare our results with the observational data in terms of the Compton $y$-parameter, and not of spectral slope.}

\end{enumerate}
To summarize, we have a set of the energy equations;
\begin{eqnarray}
 f_{\rm w} Q_{\rm tot}^{+} 
 &=& 
  y F_{\rm soft} \quad{\rm (corona)}\cr
 [1-(f_{\rm w}/2)] Q_{\rm tot}^{+}\ &=& F_{\rm soft} \quad~~{\rm (disk)} \label{energy}
\end{eqnarray}

leading to the following simple equation;
\begin{equation}
 y = \frac{f_{\rm w}}{1-(f_{\rm w}/2)},
 \label{y-fw}
 \end{equation}
which takes the maximum value of 2 when $f_{\rm w}=1$.  Note that the warm corona temperature is related to the optical depth of the warm corona, $\tau_{\rm es,w}$, through the definition of Compton-$y$ parameter,
 \begin{equation}
 kT_{\rm w} 
  = 0.20~ {\rm keV} \left(\frac{\tau_{\rm es,w}}{25}\right)^{-2} y.
\label{Tw}
\end{equation}
When $f_{\rm w} \sim 1/2$, hence, we derive $y \sim 2/3$ ($\Gamma \sim 2.4$) and
also obtain $kT_{\rm w}\sim 0.13$ keV for $\tau_{\rm es,w} \sim 25$; that is, we can reproduce the typical photon index and electron temperature of the soft excess.

This is a simple argument to demonstrate that $\Gamma \simeq 2.2 - 3.2$ and $T_{\rm w} \simeq 0.1-0.5$ keV are feasible for a moderate scattering depth,
but, what determines a coronal scattering depth?
We definitely need a more quantitative corona modeling. 
This will be done in the next section, but before that it will be instructive to check other constraints imposed for successful warm corona modeling.


\subsection{Conditions for a warm corona}
There are two key issues: 
anomalous heating in the warm corona and dominance of the Compton cooling over bound-free/free-free emission.

\subsubsection{Anomalous heating}
Some sort of {\lq}anomalous{\rq} heating mechanism should work within a warm corona.
By anomalous heating we mean that the heating rate is not simply proportional to local density nor pressure
but that it is much more enhanced within the corona; i.e.,
$f_{\rm w} \gg \Sigma_{\rm w}/\Sigma_{\rm tot} (\ll 1)$.

This is in analogue with the hot corona model by \citet{1991ApJ...380L..51H}, 
who assumed that the energy dissipation occurs not entirely within the disk body, 
but its certain fraction ($f$) is transported to coronae and is dissipated there.
The reason for this is simple (in the context of hot corona modeling); without such an anomalous heating 
the corona temperature will be too low to reproduce the observational values, $\sim 100$ keV,
because of significant Compton cooling by soft photons emerging from the disk body.
We, here, assume the same line for a warm corona \citep{2018A&A...611A..59P, 2020A&A...634A..85P}.
We will explicitly show in section 3 that anomalous heating is necessary to explain the warm corona temperature within our model.

\subsubsection{Dominance of the Compton scattering}
 What the observations indicate is that if the soft X-ray excess is produced by Comptonization in a warm corona, the cooling process in a warm corona should be dominated by Compton cooling over bound-free/free-free emissions.
Naively, this condition can be easily satisfied, 
since we know that scattering opacity ($\kappa_{\rm es}$) dominates over free-free absorption opacity ($\kappa_{\rm ff}$) 
in the inner portions of the standard-type accretion disks around black holes \citep{1973A&A....24..337S}; 
that is, the condition reads (e.g., \cite{2008bhad.book.....K}),
\begin{equation}
 r < 2.5 \times 10^3 {\dot m}^{2/3}~r_{\rm S},
\end{equation}
not depending on the black hole mass.

We should note, however, that the situation is not so simple for two main reasons:
One is that the bound-free opacity, $\kappa_{\rm bf}$, dominates in the temperature ranges well below $\sim 10^7$ K (i.e., $\kappa_{\rm abs}=\kappa_{\rm ff}+\kappa_{\rm bf}\sim \kappa_{\rm bf}$, where $\kappa_{\rm abs}$ is the absorption opacity; see Fig.2)
whereas it is not considered in the classical paper by \citet{1973A&A....24..337S}.
Roughly speaking, this leads to underestimation of the absorption opacity by a factor of 30 or so (see, e.g., \cite{1971ApJ...167..113C}).

Second, the dominance of the Compton cooling rate (over absorption cooling rate) within the corona 
is not equivalent to the condition of $\kappa_{\rm es}>\kappa_{\rm abs}$,
since we crudely estimate 
\begin{equation}
  Q^-_{\rm Comp} \sim y F_{\rm soft} \quad {\rm and} \quad
  Q^-_{\rm bf} \sim \tau_{\rm bf} \sigma T_{\rm w}^4, \label{coolingrate}
\label{cooling}
\end{equation}
where $Q^-_{\rm Comp}$ and $Q^-_{\rm bf}$ are the cooling rates due to Compton scattering and bound-free emission, respectively
(note that bound-free emission is dominant in the temperature ranges of the warm corona),
and $\tau_{\rm bf} =\kappa_{\rm bf}\Sigma_{\rm w}$
(note that $\Sigma_{\rm w}$ is defined as the surface density above the equatorial plane in the present paper).  
We assumed $\tau_{\rm bf} \ll 1$, 
since $\tau_{\rm eff} \sim 1$ and $\tau_{\rm es} \gg 1$ within the warm corona.  Note that the expression for $Q_{\rm bf}^-$ in Eq.(\ref{coolingrate}) is valid only when $\tau_{\rm bf}\ll 1$.
 From equation (\ref{cooling}) the condition of the Compton domination is 
\begin{equation}
 \frac{Q^-_{\rm Comp}}{Q^-_{\rm bf}}
　　\sim \frac{1}{\tau_{\rm bf}}
         \left(\frac{T_{\rm soft}}{T_{\rm w}}\right)^4 y > 1.
\label{dominance}
\end{equation}
We should make remark that these expressions, especially that for the absorption cooling rate in equation (\ref{cooling}), is not so accurate,
since the Kirchhoff's law is applicable only to the absorption coefficient and emissivity as functions of frequency and is not exact when we use the Rosseland mean opacity and frequency-integrated emissivity.
Nevertheless we adopt these in this study to grasp insight into the physics underlying warm corona formation.

\section{Simple model}

Let us construct a simple model based on the basic considerations presented in the previous section.
There are three basic equations: condition for the effective optical depth [equation (\ref{tau_eff})], equation of Compton $y$ [equation (\ref{Tw})], which is closely related to the energy equation [equation (\ref{energy})], and dominance of the Compton cooling [equation (\ref{dominance})]: 
\begin{equation}
 3\left(\frac{\kappa_{\rm bf}}{\kappa_{\rm es}}\right)\tau_{\rm es,w}^2=1, 
\label{basic2}
\end{equation}
\begin{equation}
T_{\rm w} = 2.4 \times 10^6~{\rm K}~y\left(\frac{\tau_{\rm es,w}}{25}\right)^{-2}, 
\label{basic1}
\end{equation}
and
\begin{equation}
 f_{\rm bf} C_{\rm w} \left(\frac{\kappa_{\rm bf}}{\kappa_{\rm es}}\right) \tau_{\rm es, w} 
    = y \left(\frac{T_{\rm soft}}{T_{\rm w}}\right)^4,
\label{basic3}
\end{equation}
where 
we set
\begin{equation} C_{\rm w} \equiv Q^-_{\rm Comp}/Q^-_{\rm bf} ~~ (>1) \quad{\rm and}\quad
  f_{\rm bf} \equiv Q^-_{\rm bf}/(\tau_{\rm bf}\sigma T_{\rm w}^4) ~~ (\sim 1)
\end{equation}
and we used the relations, $\tau_{\rm bf}=(\kappa_{\rm bf}/\kappa_{\rm es})\tau_{\rm es,w}$, $\tau_{\rm abs}\sim \tau_{\rm bf}$, and $\tau_{\rm bf} \ll \tau_{\rm es}$.  The physical meaning of the coefficient $f_{\rm bf}$ is the following: according to the Kirchhoff's law, the emissivity (per unit volume per unit frequency) divided by the opacity at a fixed frequency (per unit volume) is equal to the Planck function.  Although the frequency-integrated emissivity, $Q_{\rm bf}$, divided by the frequency-integrated opacity, $\tau_{\rm bf}$, is not necessarily equal to the frequency-integrated Planck function, i.e., $\sigma T_{\rm w}^4$, they are not so much different.  We thus suspect that $f_{\rm bf}$ should be the value with the order of unity.
The number of the unknowns is three: $T_{\rm w}$, $\tau_{\rm es,w}$, and $\kappa_{\rm bf}/\kappa_{\rm es}$. These unknowns will be uniquely determined, once we specify the value of $y$ [or $f_{\rm w}$, see equation (\ref{y-fw})].  
Note that $T_{\rm soft}$ is a function of $M$, $\dot M$, and $r$ for a given $f_{\rm w}$.

We can obtain an expression for $\tau_{\rm es,w}$ by dividing equation (\ref{basic2}) by equation (\ref{basic3}), 
\begin{equation}
 \tau_{\rm es,w} = \frac{f_{\rm bf}C_{\rm w}}{3y} \left(\frac{T_{\rm w}}{T_{\rm soft}}\right)^4.
\label{basic32}
\end{equation}
We also rewrite equation (\ref{basic1}) to have another expression for $\tau_{\rm es,w}$,
\begin{equation}
\tau_{\rm es,w} = 39 y^{1/2} \left(\frac{T_{\rm w}}{T_{0}}\right)^{-1/2},
\label{basic11}
\end{equation}
where $T_0 \equiv 10^6~{\rm K}$.
By equating equations (\ref{basic32}) and (\ref{basic11}), we finally have
\begin{equation}
 \frac{T_{\rm w}}{T_0}
     = \left[\frac{117 y^{3/2}}{f_{\rm bf}C_{\rm w}} 
            \left(\frac{T_{\rm soft}}{T_0}\right)^4\right]^{2/9}
            \simeq 2.9 \frac{y^{1/3}}{\left(f_{\rm bf}C_{\rm w}\right)^{2/9}}
            \left(\frac{T_{\rm soft}}{T_0}\right)^{8/9},
\label{simple}
\end{equation}
which implies that $T_{\rm w}$ is higher than $T_{\rm soft}$.  In addition, $T_{\rm w}$ should be lower than the hot corona temperature, $T_{\rm hot}$, as long as $y$ is on the order of unity, since the scattering optical depth of a hot corona is supposed to be the order of unity, while that of a warm corona is the order of ten from Eq.(\ref{basic11}).  We can thus understand that the warm corona temperature should be between the disk temperature ($\sim T_{\rm soft}$) and the hot corona temperature, $T_{\rm hot} \sim 10^9$ K, unless $y$ is very large and/or $f_{\rm bf}$ is very small.

We, however, notice that the derived values are somewhat underestimation.
If we set $(y, f_{\rm bf}, C_{\rm w})=(2, 1, 1)$, which maximize $T_{\rm w}$, we find $T_{\rm w}=0.47\times 10^6$ K for $T_{\rm soft}=0.1 T_0$ 
 ($=10^5$ K) and $T_{\rm w}=0.87\times 10^6$ K for $T_{\rm soft}=0.2 T_0$.
 Note that the maximum value of $y$ is 2 [which is attained for $f_{\rm w} = 1$, see equation (\ref{y-fw})].
 It may be that the correction factor could be very small, $f_{\rm bf}\ll 1$ (see discussion in the next section).

Another concern is the dependence of $T_{\rm w}$ on the black hole mass, $T_{\rm w} \propto M_{\rm BH}^{-2/9}$,
since $T_{\rm soft} \simeq T_{\rm SSdisk} \propto M_{\rm BH}^{-1/4}$ [see equation (\ref{SSdisk})].
This is inconsistent with the observations, which indicate that $T_{\rm w}$ is not so sensitive to $M_{\rm BH}$.
We should also remark that $T_{\rm w}$ should rapidly decrease outward; $T_{\rm w} \propto r^{-2/3}$,
since $T_{\rm SSdisk}\propto r^{-3/4}$ [see equation (\ref{SSdisk})]. 
This leads to the conclusion that warm corona should exist only in the black hole vicinity.

\section{Discussion}
In this section, we first compare our model with the similar study on a warm corona by \citet{2015A&A...580A..77R}, discussing the similarities and differences between them.  In the following subsections, we discuss the thermal structure of a warm corona taking into account both bound-free cooling and Compton cooling, the anomalous heating in a warm corona, and the dependence of the temperature of a warm corona on its parameters. 

\subsection{Comparison with \citet{2015A&A...580A..77R} and \citet{2020A&A...634A..85P}}
Let us compare our model with the previous studies by \citet{2015A&A...580A..77R} and \citet{2020A&A...634A..85P} on a warm corona as the origin of soft X-ray excess.  \citet{2015A&A...580A..77R} considered a purely scattering layer as a warm corona, ignoring emission/absorption, while we carefully consider the effects of bound-free cooling in it.  We define the height of the warm corona in terms of the effective optical depth (which should be around unity), while they calculated the height of the warm corona from the hydrostatic balance.  A big distinctive consequence which arises by considering emission/absorption is that optical depth is no longer simply proportional to surface density ($\Sigma$) but their relationship depends also on density and temperature ($\tau=\kappa(\rho,T)\Sigma/2$, in general).  We thus need three conditions for the warm corona, Eqs. (\ref{tau_eff}), (\ref{Tw}), and (\ref{dominance}), to fix its structure.

The arguments in \citet{2020A&A...634A..85P} are two-fold: first, they adopted the same condition (of pure scattering), and second they used the existing codes of radiation transfer, in which the bound-free emissions are incorporated.  They showed their results in the two-dimensional plane (i.e., as functions of two parameters, the coronal optical depth and the fraction of energy dissipated in the warm corona).  By contrast, we fix the physical quantities, such as corona temperature, density, optical depth, and so on, as functions of $M_{\rm BH}$, $\dot{M}$, and $r$ from the three conditions.  Probably, our solutions could be in their 2-dimensional plots, but it is left as future work to directly compare our model and theirs.

To summarize, although the motivation of this study (construction of simple semi-analytical models for a warm corona) is the same as that of \citet{2015A&A...580A..77R} and \citet{2020A&A...634A..85P}, and the approach looks similar, but the adopted basic conditions are distinct.

\subsection{Bound-free cooling vs. Compton cooling}
\subsubsection{Bound-free cooling curve} 
\label{cooling curve}
\begin{figure}[ht]
 \centering\includegraphics[width=12.cm]{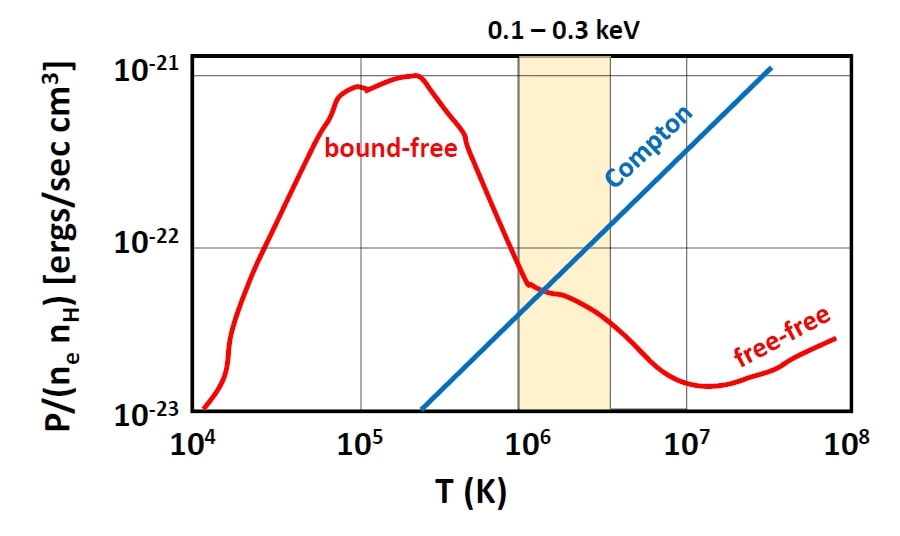}
 \caption{Schematic view of cooling curves: the bound-free/free-free cooling rate (after Cox \& Daltabuit 1971) and the Compton cooling rate.
}
 \label{fig:cooling}
 \end{figure}

One of the largest uncertainties in the simple model presented in the previous section is the crude evaluation of the bound-free cooling rate [equation (\ref{cooling})].
For reference, we show in figure \ref{fig:cooling} a schematic diagram of the bound-free cooling rate of plasma in collisional ionization equilibrium as a function of temperature 
(adapted from Fig.1 of \cite{1971ApJ...167..113C}). 
Contributions by heavy elements, such as O, Ne, Mg, Si, and S, are dominant in the temperature range around $\sim 10^6$ K (see also \cite{1969ApJ...157..1157C}).

 Remarkably, the bound-free cooling curve shows a rapid decrease with increasing temperature in the temperature range below several million degree K (which we are concerned with). Roughly, we find $q^-_{\rm bf} \propto T^{-2}$ (or $\propto T^{-0.5}$) below (above) $\sim 10^6$ K, where $q^-_{\rm bf}$ represents a bound-free cooling rate per unit volume, per electron, and per ion.  This is due to the collisional ionization of metal elements.

In reality, however, we need non-grey calculations. Also needed are the considerations of photo-ionization effects, which are totally missing in Figure (\ref{fig:cooling}).
More accurate evaluations are needed as future work in this point, but we wish to emphasize that our results will not be so sensitive to the precise treatments of the bound-free cooling rate, since it is Compton cooling, and not bound-free cooling, that is balanced with anomalous heating (see below).

\subsubsection{On the dominance of Compton cooling} 
In figure \ref{fig:cooling} we also schematically plot the Compton cooling rate, which is an increasing function of temperature, as long as we assume that $\Sigma_{\rm w}$ is kept constant.  Note that its magnitude is taken arbitrarily, since it varies, depending on $\Sigma_{\rm w}$ (or $H_{\rm w}=\Sigma_{\rm w}/\rho_{\rm w}$, where $\rho_{\rm w}$ is the mass density of the warm corona) and $F_{\rm soft}$.
We thus understand that the bound-free cooling rate (or the Compton cooling rate) exhibits negative (or positive) temperature dependence (see next subsection).

In our simple model, we introduced a parameter, $C_{\rm w}$ (ratio of the Compton cooling rate to the bound-free cooling rate in the warm corona), 
whose exact value cannot be determined within our model. This should be greater than unity in order to account for the observations, and this condition can be fulfilled by adjusting the height of the warm corona ($H_{\rm w}$), keeping its surface density $\Sigma_{\rm w}$ constant. 
Actually, the Compton cooling rate ($Q_{\rm Comp}^- \propto \tau_{\rm es,w}^2\propto \Sigma_{\rm w}^2$) remains unchanged, as long as $\Sigma_{\rm w}$ is kept constant, while the bound-free cooling rate is
\begin{equation}
\label{qrad}
  Q_{\rm bf}^-\propto q_{\rm bf}^-\rho_{\rm w}^2H_{\rm w} \propto q_{\rm bf}^-\Sigma_{\rm w}^2/H_{\rm w}
\end{equation}
is inversely proportional to $H_{\rm w}$.  From the condition of $C_{\rm w}>1$, one can derive the condition for $H_{\rm w}$ as
\begin{eqnarray}
H_{\rm w} > \frac{4k_{\rm B}T_{\rm w}}{m_e c^2}F_{\rm soft}\kappa_{\rm es}^2 q_{\rm bf}^{-2}.
\end{eqnarray}
The height of the warm corona, in turn, depends on how much anomalous heating takes place at which height, and it depends on the detailed magnetohydrodynamic (MHD) processes, such as magnetic reconnection, within the corona.  This is very difficult to evaluate and we need await future radiation-MHD simulations with high numerical resolutions (see discussion in sec. \ref{multi-zone}).

\subsubsection{Thermal stability: simple argument}

 \begin{figure}[ht]
 \centering\includegraphics[width=9.cm]{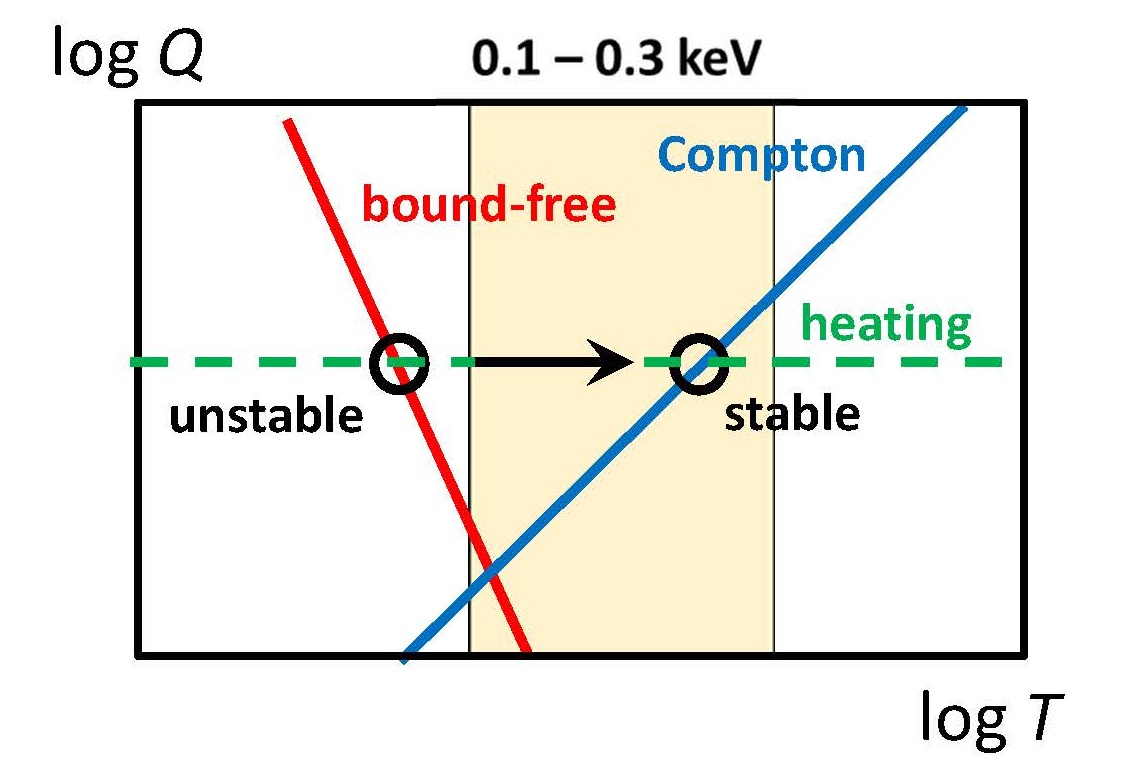}
 \caption{Schematic figure explaining the relationship between the shape of the heating and cooling curves and the thermal stability.}
 \label{fig:stability}
 \end{figure}

 The different temperature dependence of the radiative and Compton cooling rates may lead to an interesting consequence. 
Let us examine the stability of the warm corona exposed to constant soft radiation flux from the disk body ($F_{\rm soft}$) 
for given surface density ($\Sigma_{\rm w}$) and heating rate.  Figure \ref{fig:stability} is a schematic picture drawing the bound-free cooling rate, the Compton cooling rate, 
and the heating rate as functions of temperature (all the heating/cooling rates are measured per unit surface area).
We understand that the bound-free cooling rate rapidly drops as temperature increases because of the ionization (see equation \ref{qrad}).  By contrast, the Compton cooling rate is an increasing function of temperature, when $\Sigma_{\rm w}$ and thus $\tau_{\rm w}$ are fixed.
By assumption, we fix the value of the heating rate ($Q^+$).\footnote{We here assume only one simple case, since we are not aware of the precise functional form of the anomalous heating rate. The precise shape of the heating curve will not affect the following discussion, unless the slope of the heating curve is greater than $\propto T$ or less than $Q^+\propto T^{-2}$.}

We just point out that 
there could be two thermal equilibrium solutions, in which heating rate is equal to cooling rate. 
One is the bound-free branch (in which the bound-free emission dominates) found at lower temperatures
and another is the Compton branch (in which Compton-cooling dominates) found at higher temperatures.
From Fig. \ref{fig:stability}, we understand that not only the Compton cooling rate but also the heating rate
should exceed the absorption cooling rate at $T\sim (1-3)\times 10^6$   K in order for a warm corona to appear.  In other words, the occurrence of anomalous heating should be necessary, since otherwise the heating rate would be below the intersection of the two cooling curves.  Since the cooling rate is the sum of the bound-free cooling rate and the Compton cooling rate, there would be no thermal equilibrium solution in such a case.

Suppose the situation in general that there are multiple thermal equilibrium solutions, which one will be realized?
The answer is, the one(s) which is (are) thermally stable.
We thus need to check the criterion for the thermal stability; that is 
\begin{equation}
   \left(\frac{\partial Q^+}{\partial T}\right)_{\Sigma_{\rm w}} <
    \left(\frac{\partial Q^-}{\partial T}\right)_{\Sigma_{\rm w}},
\label{stability}
\end{equation}
where $Q^+$ and $Q^-$ are the heating and cooling rate per unit surface area
and differentiation is made at constant $\Sigma_{\rm w}$, 
since thermal timescale is much shorter than the viscous timescale, over which $\Sigma_{\rm w}$ varies
(see, e.g., \cite{1975PThPh..54..706S, 1976MNRAS.175..613S, 1981ARA&A..19..137P};
see also Chapter 4 of \cite{2008bhad.book.....K} for more generalized discussion).

Let us examine the two thermal equilibrium solutions found in Fig. \ref{fig:stability}. 
We can immediately understand that the Compton branch is thermally stable,
while the bound-free branch is thermally unstable.  In conclusion, it is natural that the observations of soft X-ray excess show the Compton-dominated spectra,
which are very smooth without any atomic feature.

What exactly happens in the warm corona?
Suppose that the disk surface layer (with $\tau_{\rm eff} \sim 1$) is on the bound-free branch.
(The effective optical depth is assumed to be unity there.)
Since such a layer is thermally unstable, its temperature should either decrease or increase.  When the temperature decreases down to the surface temperature of the disk body; 
that is, the warm corona will disappear.  When the temperature increases, conversely, a transition to the thermally stable Compton branch will be completed.
We should note, however, the effective optical depth is no longer unity but should decrease 
as a result of decrease of $\kappa_{\rm bf}$ at higher temperatures.
This may indicate that the warm corona may have a multiple-zone structure (see discussion in sec. \ref{multi-zone}).

\subsubsection{Thermal stability: comment}
We need to make a remark on the stability analysis made by \citet{2003A&A...412..317C}
(see also similar but independent discussion by \cite{1993PASJ...45...775}).
They examined the thermal stability of the disk-corona system starting with the similar energy equation to our equations (\ref{disk-heating}) and (\ref{disk-cooling}), but explicitly write down the expression for the energy transportation rate, 
and find the marginal stability condition as $f_{\rm w}  = 0.5$ ($w = 0.5$ in their notation).

At first glance, our result seems an apparent contradiction with theirs, but this is not the case, since the situation is different.
They examined the stability of the disk body, assuming that a certain fraction of the energy produced in the disk body is transported to a corona, and examined the condition how the original criterion is affected. 
[As is well known, the standard-type disk suffers thermal and viscous instabilities, when radiation pressure dominates over gas pressure \citep{1974ApJ...187L...1L,1975PThPh..54..706S,1976MNRAS.175..613S}.]
By contrast, we examined the stability of the corona, assuming that the disk body is stable (and hence the radiation flux from the disk body are unchanged).
The case of thermally unstable disk body need to be examined as future work. 

\subsection{Anomalous heating}
\subsubsection{On the value of $f_{\rm w}$}
As shown in Sec. 2, the fraction of the accretion energy that is dissipated in a warm corona, $f_{\rm w}$, should be of order unity to reproduce the observational features of soft X-ray excess within our model, since otherwise the temperature and so the Compton-$y$ parameter of a warm corona would be much less, which means that the warm Comptonized emission would be too weak to observe. 
Note that the effects of internal Compton scattering on the structure and spectra of scattering-dominated accretion disk (with no anomalous heating) were intensively discussed in the 1980's and 1990's (see, e.g., \cite{1987ApJ...321..305C, 1992MNRASA...258...189, 1995ApJ...440..610S}). 
Although Compton up-scattering produces enhancements in the Wien part of the UV bump emission, the enhancements are not enough to explain the soft excess component (without anomalous heating).  Some previous studies (e.g., \cite{2013A&A...549A..73P,2020A&A...634A..85P}) also introduced the parameter $f_{\rm w}$ in the models and constrained its value as being close to unity.  The difference between them and our study is that we put the condition for a warm corona whose effective optical depth is unity, while the previous studies did not considered such a condition.  This makes some quantitatively different results while the condition of $f_{\rm w}\simeq 1$ is common.

The magnetic energy dissipation in the upper layer of an accretion disk in AGNs or X-ray binaries has been discussed previously to reproduced their observational features \citep{2000MNRAS.318L..15M,2015ApJ...809..118B,2020A&A...633A..35G,2023A&A...675A.198G}.  Interestingly, the significant energy dissipation at the disk surface layer has been implied by several numerical simulations of an accretion disk \citep{2003ApJ...593..992T, 2006ApJ...640..901H, 2009ApJ...707..428B, 2011ApJ...736....2O,2018ApJ...857...34Z,2019ApJ...885..144J,2020MNRAS.492.1855M}. 
They performed the multi-dimensional radiation-magnetohydrodynamical simulations of black hole accretion flows and outflows for various accretion rates, and presented their global structures.  Especially they have shown the vertical dependence of the energy dissipation rate per unit volume and magnetic energy density becomes shallower as the distance from the equatorial plane increases.  This means that it is not necessary for the accretion energy to dissipate mainly at the equatorial plane of the disk, and that a significant heating may occur at the disk surface layer.  More accurate estimation regarding the heating rate distribution in a scattering-dominated layer is needed as future work.

\subsubsection{Limitation of the one-zone model}
\label{multi-zone}
The big assumption made in our simple analysis resides in (vertically) one-zone approximation.
It is more likely that the anomalous heating occurs in a non-uniform fashion
so that a warm corona may have a multi-zone structure.
Suppose that the anomalous heating takes place predominantly only in an upper layer of the warm corona,
while heating is much less in other parts.
It then follows that the optical depth of the heated layer with a temperature of $\sim T_{\rm w}$ could be less than unity.
\footnote{We still keep the terminology of the warm corona to be a layer with $\tau_{\rm eff} \sim 1$ formed above the disk body.
Note also that the optical depth of the unheated region cannot be negligibly small, since the absorption opacity is normally a decreasing function of temperature so that the value of $\kappa_{\rm bf}$ is larger in the unheated, high-density region than in the heated, low-density one.}

Then, we may have an inequality,
\begin{equation}
 3\left(\frac{\kappa_{\rm bf}}{\kappa_{\rm es}}\right)\tau_{\rm es,w}^2\equiv\tau_{\rm eff}^2 < 1, 
\label{basic22}
\end{equation}
[instead of equation (\ref{basic2})], leading to
\begin{equation}
 \frac{T_{\rm w}}{T_0}
            \simeq 2.9 \frac{y^{1/3}}{\left(f_{\rm bf}C_{\rm w}\tau_{\rm eff}^2\right)^{2/9}}
            \left(\frac{T_{\rm soft}}{T_0}\right)^{8/9},
\label{simple2}
\end{equation}
[instead of equation (\ref{simple})].
We can then find a higher temperature solution for $\tau_{\rm eff} < 1$.
If we set $(y, f_{\rm bf}, C_{\rm w}, \tau_{\rm eff})=(2, 1, 1, 0.1)$, for example, we find $T_{\rm w}=1.3\times 10^6$ K for $T_{\rm soft}=0.1 T_0$ ($=10^5$ K).
In conclusion, partial (surface) heating of an upper layer ($\tau_{\rm eff}<1$ but $\tau_{\rm es}\gg 1$) of the warm corona is probably a solution for making the warm corona temperature above $\sim 10^6$ K.

\subsubsection{Difference between a warm corona and a hot corona}
Both a warm corona and a hot corona are introduced to theoretical models of AGNs as Comptonizing plasma that lie above the body of an accretion disk, which provide thermal seed photons to be Compton up-scattered.  
Their origins are distinct, however.  

A warm corona is introduced in our model as a natural extension of accretion disk models; that is a scattering-dominated layer at the surface.  In this sense, a warm corona can exist whenever or wherever there is a scattering-dominated atmosphere above an optically-thick accretion disk. A noteworthy feature of the warm corona is that the temperature difference between the disk and corona is not so large ($T_{\rm w}\sim 10^6~{\rm K}$ and $T_{\rm SSdisk}\sim 10^5~{\rm K}$) that the thermal conduction flux between them cannot be effective.  

In the case of the hot corona, by contrast, the temperature difference should be much larger
($T_{\rm hot}\gtrsim 10^9~{\rm K}$ in the corona), which is required to account for the hard power-law component in the X-ray spectra of AGNs. Therefore, the heat conduction from a hot corona to the underlying disk is expected (as in the case of solar corona), and it can drive the mass evaporation from a disk into a corona (\cite{1995A&A...300..823L, 1999ApJ...527L..17L, 2000A&A...361..175M, 2002ApJ...575..117L}).  This is a big distinction between a warm corona and a hot corona.  How these two components coexist in the innermost part of an accretion flow is beyond the scope of this work.

Let us consider what will occur when anomalous heating takes place at scattering-dominated atmosphere.  The atmospheric temperature should increase, when an anomalous heating sets in, but the temperature increase will be up to $\sim 10^6$ K, at which Thomson thick Compton cooling is balanced with anomalous heating, as we have so far discussed. The temperature increase is moderate (i.e., $T_{\rm w}\ll 10^9~{\rm K}$)
to acheive the condition of Compton $y \sim 1$ when $\tau_{\rm es}\gg 1$ and $\tau_{\rm eff}=1$.  Note that $T_{\rm w}$ can be higher if the condition of $\tau_{\rm eff}$ is relaxed to allow that it can be smaller than unity.

\subsection{Parameter dependence}

\subsubsection{Dependence on the accretion rate }

 \begin{figure}[ht]
 \centering\includegraphics[width=9.cm]{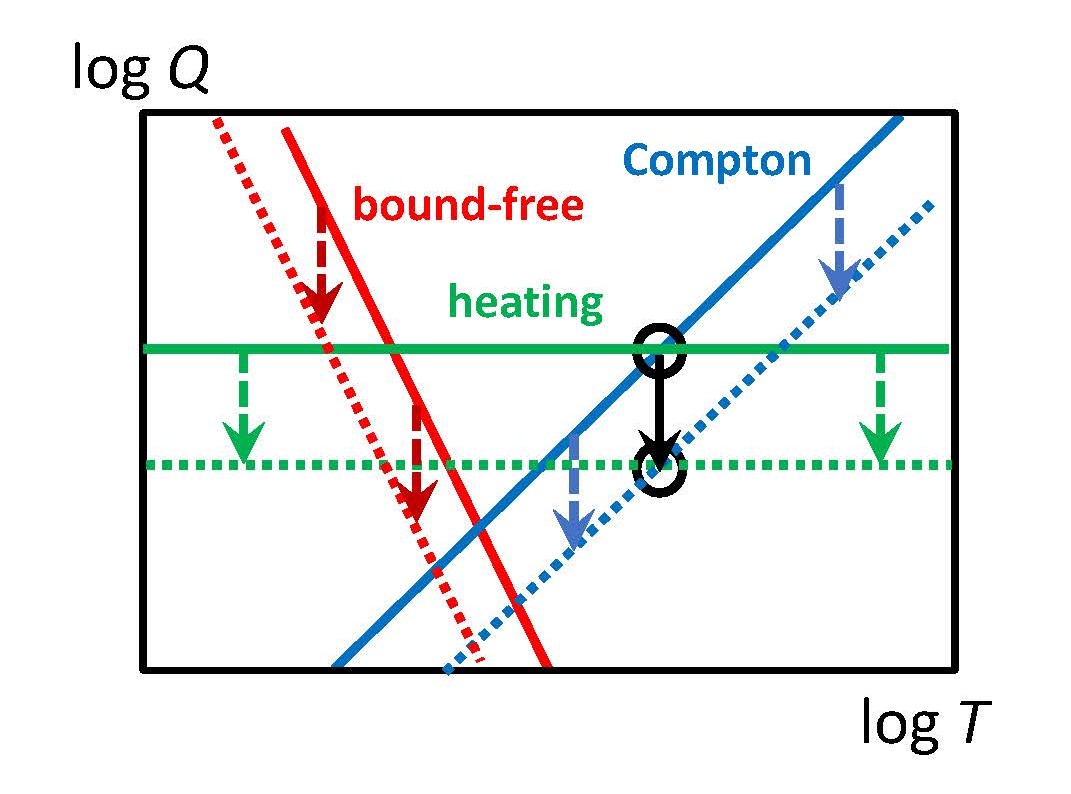}
 \caption{Schematic figure explaining how the heating and cooling curves shift, when accretion rate ($\dot M$) decreases.
 The solid (or dotted) lines represent the heating/cooling rates before (after) decreasing $\dot M$.}
 \label{fig:mdot-dep}
 \end{figure}

It is observationally indicated that the warm corona temperature $T_{\rm w}$ is rather insensitive to the black hole mass or luminosity (or equivalently, accretion rate).
How can we explain these facts?

We first discuss the accretion-rate dependence on $T_{\rm w}$.
Suppose that the warm corona lies on the Compton branch as is displayed in figure \ref{fig:stability} and let mass accretion rate decrease for a fixed $M_{\rm BH}$ and fixed $r$.
We immediately understand that the coronal heating rate decreases for a fixed $f_{\rm w}$, since it is proportional to $\dot M$ [see equations (\ref{corona-heating}) and (\ref{Qtot})].
We can also easily understand that the Compton cooling rate should also decrease in proportion to $\dot M$ for a fixed $\Sigma_{\rm w}$, since $F_{\rm soft} \sim F_{\rm SSdisk} \sim Q_{\rm tot}^+$, as long as $f_{\rm w} = O(1)$ [see equation (\ref{energy})].
Since both rates decrease in the same way, the equilibrium temperature should be kept the same (see figure \ref{fig:mdot-dep}).
This naturally explains why $T_{\rm w}$ is insensitive to $\dot M$.

Note that the bound-free cooling rate should also decrease, since $Q_{\rm bf}^- \propto \rho^2 H_{\rm w} \approx \rho\Sigma_{\rm w}$
and $\rho$ decreases, as $\dot M$ decreases. Although the amount of decreases may not be the same as that of the heating (or Compton cooling) rate, this is not critical, since the equilibrium temperature is determined not by the bound-free rate but by the Compton rate.

Note also that there is a lower limit to the mass accretion rate, for which our model can apply.
When the accretion rate is so small, sufficient amount of material can no longer be supplied to the surface layer to fulfill the condition of $\tau_{\rm eff} \sim 1$.  The exact expression for this lower limit is difficult in the framework of the one-zone model.  We leave this issue to future works.

\subsubsection{Dependence on the black hole mass}
The next issue is how to understand that $T_{\rm w}$ does not depend on $M_{\rm BH}$.  In the framework of the simple model presented in Sec. 3, $T_{\rm w}$ should have $M_{\rm BH}$ dependence.  One promising way to overcome this issue is to introduce a multi-zone structure for the warm corona (or, equivalently, to relax the condition of $\tau_{\rm eff} = 1$).
 We again use figure \ref{fig:mdot-dep}.
Suppose, this time, that $M_{\rm BH}$ increases, while keeping ${\dot m} (\equiv {\dot M}c^2/L_{\rm Edd})$ constant, at the same normalized radius ($r/r_{\rm S}$).
Since $Q_{\rm tot}^+ \propto M_{\rm BH} {\dot M}/r^3 \propto {\dot m} (r/r_{\rm S})^{-3} M_{\rm BH}^{-1}$,
we understand that the heating rate should decrease in inversely proportional to $M_{\rm BH}$.
So does Compton cooling rate.
Since the heating and Compton cooling rate decreases in the same fashion, the equilibrium temperature should not change, as in the case of changing $\dot M$.
The bound-free cooling should also decrease, since it is proportional to $\rho$ and $\rho$ usually decreases with increase of $M_{\rm BH}$ for a fixed $\dot m$ and $r/r_{\rm S}$. 
In fact, recalling the relationship, ${\dot M} = - 2 \pi r v_r \rho H \propto r^2 \rho$ (as long as $v_r$ and $H/r$ are not so sensitive to $M_{\rm BH}$), one can derive $\rho \propto M_{\rm BH}^{-1}$ for a fixed $\dot m$ and fixed $r/r_{\rm s}$.  Its precise $M_{\rm BH}$ dependence does not matter, since it does not affect the equilibrium temperature.
We thus understand why $T_{\rm w}$ is not scaled by $M_{\rm BH}$.

We should note, however, that there is a lower limit to the black hole mass, since $T_{\rm soft}$ increases with a decrease of $M_{\rm BH}$ and eventually exceeds the typical warm corona temperature of $\sim 10^6$ K.
Thus, the warm corona solution disappears in small $M_{\rm BH}$ objects.

As for the strength of the soft X-ray excess, it has also been indicated to be insensitive to the black hole mass \citep{2004MNRAS.349L...7G, 2022arXiv221011977M}.  In our model, the soft X-ray excess strength is determined by the energy dissipation fraction, $f_{\rm w}$, which is assumed as a free parameter in this work.  In order to interpret this observational fact within our warm corona scenario, it is necessary to model how $f_{\rm w}$ should be determined from the physical properties of an accretion disk such as $M$ or $\dot{M}$, which is also left as a future work.

\subsubsection{Dependence on the radius}
The next issue is the radial dependence of $T_{\rm w}$.
We have suggested in section 3 that $T_{\rm w}$ may be a strong function of radius $r$, since it is roughly scaled with $T_{\rm soft} \approx T_{\rm SSdisk} \propto r^{-3/4}$.
In other words, a warm corona can exist only at small radii.
This can be easily understood, since much cooler seed photons emerging at larger radii requires a large Compton amplification factor, or large $y$, in contradiction with the observation (which indicate similar photon indices and so similar $y$-values). 

\section{Concluding remarks}
\begin{table}[h]
\caption{Comparisons between hot corona, warm corona, and super-Eddington corona.}
\begin{tabular}{llll}
\hline
& hot corona & warm corona & super-Eddington corona	\\
\hline
Temperature	& $\sim 10^9$ K & $\sim 10^6$ K & a few $\times 10^7$ K	\\
(Photon energy)	& ($\sim 100$ keV) & ($\sim 0.1$ keV) & (a few keV)		\\
Compton $y$ & $\sim 1$ & $\sim 1$ & $\sim 1$			\\
Thomson $\tau$ & $\sim 1$ & $\sim 30$ & $\sim 10$			\\
Effective $\tau$ & $\ll 1$ & $\sim 1$ (or less) & $\ll 1$ 			\\
Where? & ubiquitous & only around massive BHs & only above super-Eddington flow	 \\
Origin (?) & evaporation of disk gas & anomalous heating & radiation-pressure driven outflow	 \\
\hline
\end{tabular}
\end{table}

In this study we attempt to construct a vertically one-zone model of a warm corona as the origin of the soft X-ray excess component found in the X-ray spectra of AGNs.  In this model, the soft X-ray excess is accounted for as the unsaturated Compton scattering in the scattering-dominated layer (with an effective optical depth of $\tau_{\rm eff}\lesssim 1$) above and below an accretion disk body. 
Assuming that a significant fraction of accretion energy is dissipated within this layer, and that the Compton cooling dominates the bound-free cooling there, it can act as a warm corona that has an intermediate temperature between the disk ($\sim 10^5~{\rm K}$) and the hot corona ($\sim 10^9~{\rm K}$).  In the previous studies on a warm corona, the warm corona is considered as a separate component formed above the top of an accretion flow, while this study has shown that it is naturally formed within the surface layer of an accretion flow above the photosphere where the effective optical depth from infinity is $\tau_{\rm eff}=1$.
In table 1 we compare distinctive nature of different types of corona: hot corona, warm corona, and super-Eddington corona (which is formed above a super-Eddington accretion flow; see e.g., \cite{2021PASJ...73...630K}).

In Introduction, we addressed a number of issues regarding the warm corona model.
Here, we examine which ones can be answered and which ones are left as future work.
\begin{itemize}
\item {\bf Why similar coronal temperatures regardless of $M_{\rm BH}$ or $\dot M$?}

This is probably because both of the anomalous heating rate and Compton cooling rate are proportional to the dissipation rate of the accretion energy, and so the equilibrium temperature would not change, even if $M_{\rm BH}$ and/or $\dot M$ would vary.

Precisely speaking, however, the derived $T_{\rm w}$ values are somewhat underestimation.
The reason for this is not yet clear, but a promising possibility is that it may reflect the limitation of the one-zone model.
If only a surface of the layer with $\tau_{\rm eff} \sim 1$ would be heated up (i.e., if we would relax the condition to be $\tau_{\rm eff} < 1$), the temperature ($T_{\rm w}$) could be higher. Construction of a multi-layer model is left as a future issue.  Specification of the place and magnitude of anomalous heating is another issue to be investigated by means of multi-dimensional radiation-MHD simulations.    We note that In this scenario, \citep{2020A&A...634A..85P} and  \citep{2023A&A...675A.198G} claimed that Compton cooling is dominant in a warm corona if the anomalous heating is large enough.  We take a similar approach, but the nature of a warm corona is distinct.

\item {\bf Why similar photon indices?}

Since similar photon indices mean similar Compton $y$-parameter in the framework of the warm corona model [see equation (\ref{Gamma})],
this question is rephrased as {\lq\lq}why is the $y$-parameter always around unity?{\rq\rq}.
This is because the photons lastly undergo substantial energy change due to Compton up-scattering at the layer with 
$y = 1$ \citep{1987ApJ...321..305C} before reaching an observer.
Soft photons entering the lower warm corona undergo multiple Compton scattering
and lose energy by each Compton scattering. (That is, they lose their original memory.)
The observed spectra are thus formed within the upper warm corona with $y=1$.

\item {\bf Why not observed in stellar-mass black holes?}

This is because soft photon energy from the disk around stellar-mass black holes are much higher,
$T_{\rm soft} \sim 10^7$ K.
In order to produce soft-excess photons with $0.1 - 0.3~{\rm keV}$ via Compton up-scattering,
the energy of injected soft photons should be less than $kT_{\rm soft} < 0.1$ keV 
(or $T_{\rm soft} < 10^6$ K). From the standard-disk relation [equation (\ref{SSdisk})], however, we find $T_{\rm SSdisk}\simeq 3\times 10^6$ K at $r = 10~r_{\rm S}$ for $M\sim 10 M_\odot$. 
We thus conclude that $M$ should be greater than $\sim 10^3~ M_\odot$, at least, to have $T_{\rm soft} < 10^6$ K.  It may be possible, in principle, to make a corona with temperature between $10^7~{\rm K}$ and $10^9~{\rm K}$, but it is very difficult in the framework of our model, since at such high temperatures the bound-free cooling will drop significantly so that it will be extremely difficult to achieve the condition of $\tau_{\rm eff} \simeq (3\tau_{\rm bf} \tau_{\rm es})^{1/2} = 1$
(Note that smaller $\tau_{\rm bf}$ means larger $\tau_{\rm es}$ from this relationship, and large $\tau_{\rm es}$ requires that $T_{\rm w}$ should be very low from the condition of $y~(\propto T_{\rm w}\tau_{\rm es}^2)=1$, in contradiction to the assumption of higher temperatures).

\item {\bf What determines the strengths of the soft excess?}

In our model, the soft X-ray excess strength is determined by the energy dissipation fraction, $f_{\rm w}$, which is an unknown parameter.
We need finer-resolution radiation-magnetohydrodynamical simulations, in combination with radiation transfer calculations incorporating Compton scattering, to calculate the energy dissipation rate as a function of the vertical height and the emergent spectra.

\item {\bf What is the geometrical location of the warm corona?}

Since the conditions for producing a warm corona are much severer than those for a hot corona because its temperature sensitively depends on radius (see Eq.\ref{simple}), our naive expectation is that only the innermost part of the disk is covered by a warm corona. 
Again, we need further study to settle down this issue.

\item{\bf Why is no corona with intermediate temperatures of 1 - 10 keV observed}

This is because we only have two thermally stable solutions for coronae on the condition of the existence of anomalous heating: hot corona solutions (with $\tau_{\rm es} \sim 1$ and $kT_{\rm hot} \sim 10^2$ keV)
and warm corona solutions (with $\tau_{\rm eff} \sim 1$ and $kT_{\rm w}\sim 0.1$ keV). 

\end{itemize}

We are grateful to Hajime Inoue, Ryoji Matsumoto, Chris Done, Ken Ohsuga, Tohru Nagao, Aya Kubota, Hirofumi Noda, and Misaki Mizumoto for their valuable comments.  This work is supported in part by JSPS KAKENHI Grant Number 22K03686 (N.K.) and 20K34567 (S.M.).


\end{document}